\documentclass[11pt,a4paper]{article}

\usepackage[margin=1in]{geometry}
\usepackage{graphicx}
\usepackage{booktabs}
\usepackage{amsmath}
\usepackage{hyperref}
\usepackage{caption}
\usepackage{subcaption}
\usepackage{float}
\usepackage{parskip}
\usepackage{titlesec}
\usepackage{xcolor}
\usepackage{microtype}
\usepackage{enumitem}
\usepackage{abstract}

\hypersetup{
    colorlinks=true,
    linkcolor=blue!60!black,
    citecolor=blue!60!black,
    urlcolor=blue!60!black,
    pdftitle={Comparing Retrieval Methods for Academic Advisor Discovery},
    pdfauthor={Biraj Subedi},
    pdfsubject={Information Retrieval},
    pdfkeywords={expertise retrieval, faculty search, semantic embedding,
                 BM25, learning-to-rank, academic advisor discovery}
}

\titleformat{\section}{\large\bfseries}{{\thesection.}}{0.5em}{}
\titleformat{\subsection}{\normalsize\bfseries}{{\thesubsection}}{0.5em}{}

\title{
    \vspace{-1cm}
    {\LARGE \textbf{Comparing Retrieval Methods for Academic Advisor Discovery}} \\[0.4em]
    {\large A Six-Method Study of 768 CS Faculty Profiles}
}
\author{
    Biraj Subedi \\
    \small Independent Researcher \\
    \small \href{https://github.com/subedibiraj/academic-discovery}{github.com/subedibiraj/academic-discovery}
}
\date{July 2026}

\begin{document}
\maketitle

\begin{abstract}
We present a comparative evaluation of six information retrieval methods
for the task of academic advisor discovery: ranking CS faculty members
by relevance to a graduate applicant's research interest statement.
The methods span sparse lexical matching (Jaccard overlap, TF-IDF, BM25),
dense semantic retrieval (\texttt{all-MiniLM-L6-v2} sentence embeddings),
hybrid score fusion, and learning-to-rank.
Evaluation uses a new domain-specific collection: 768 faculty profiles
scraped from 9 US CS departments, with 162 graded relevance judgments
(grade 0/1/2) across 5 queries representing distinct graduate student
research profiles.

Across all five queries, \textbf{Reranked achieves the highest mean
NDCG@10 (0.477, std 0.138)}, followed by Semantic (0.450), Hybrid (0.421),
BM25 (0.406), Jaccard (0.303), and TF-IDF (0.246).
After Bonferroni correction across all 15 pairwise comparisons
($\alpha = 0.05/15 = 0.0033$), TF-IDF is significantly worse than
BM25, Semantic, Hybrid, and Reranked ($p_{\text{Bonf}} < 0.001$ for all four);
no other pairwise difference survives correction at 5 queries.
BM25 is the most consistent method (std = 0.090 across queries),
making it the most reliable choice when query domain is unknown.

A field ablation (Q1, 67 labels) reveals that biography alone (NDCG 0.634)
\emph{outperforms} the full model combining biography with research area
tags (0.593): research area tags act as noise when prepended to biography,
reducing NDCG by 0.040. Removing biography entirely drops NDCG by 0.130.
A controlled experiment on a 296-professor subpopulation shows that
concatenating arXiv paper abstracts \emph{reduces} NDCG@10 by 0.176
(0.772~$\to$~0.596); relevant professors suffer larger average score drops
than irrelevant ones ($-$0.238 vs.\ $-$0.139, $n=3$ relevant professors),
suggestively consistent with a jargon-dilution explanation and motivating
a late-fusion architecture.
Learning-to-rank (67 labels) achieves higher MAP@10 than Semantic
(0.480 vs.\ 0.388) but lower NDCG@10 (0.568 vs.\ 0.593); neither
difference is statistically significant.

All code, scrapers, and relevance labels are released openly.\footnote{%
\url{https://github.com/subedibiraj/academic-discovery}}
\end{abstract}

\vspace{0.5em}
\hrule
\vspace{1em}

\section{Introduction}

\subsection{The Problem}

Identifying a suitable research advisor is a central challenge for
graduate school applicants, yet it receives little attention as an
information retrieval task.
The standard approach is manual: an applicant browses faculty profile
pages across target universities, reads research interest statements,
scans recent papers, and judges fit --- repeating this for potentially
hundreds of profiles across 8--10 institutions.

This process has three structural limitations.
First, \emph{vocabulary mismatch}: a student interested in
\textit{``extracting structured information from web data''} and a
professor who studies \textit{``knowledge discovery from heterogeneous
networks''} are closely aligned, but keyword search returns nothing
because the surface terms do not overlap.
This is exactly the problem that dense semantic retrieval was designed
to address~\cite{karpukhin2020dpr,reimers2019}.
Second, \emph{no ranked output}: manual browsing produces no relevance
ranking, forcing the applicant to hold all comparisons in memory.
Third, \emph{information fragmentation}: research interests, biography,
publications, and contact details are spread across inconsistent page
structures at different universities, making systematic comparison
difficult.

Despite the clear IR framing of this task, no prior work has
benchmarked retrieval methods specifically for faculty advisor discovery,
nor constructed a relevance-judged evaluation collection for it.
Existing academic search systems such as AMiner~\cite{tang2008arnetminer}
and WISER~\cite{cifariello2019wiser} target general academic expert
finding across large corpora, rather than the targeted, applicant-centric
retrieval task we study here.

\subsection{Our Approach and Contributions}

This paper makes three contributions.

\textbf{(1) A new evaluation collection.}
We construct a domain-specific IR benchmark for faculty advisor retrieval:
768 faculty profiles from 9 US CS departments scraped and normalized into
a unified schema, with 162 graded relevance judgments across 5 queries
representing distinct graduate student research profiles.
No comparable collection exists for this task.
All profiles, labels, and evaluation scripts are released openly.

\textbf{(2) A systematic retrieval comparison.}
We benchmark six methods spanning the lexical-to-semantic spectrum ---
Jaccard overlap, TF-IDF, BM25, sentence-embedding similarity, hybrid
score fusion, and learning-to-rank --- using standard IR evaluation
metrics (NDCG@10, MAP@10, Precision@10) with bootstrap significance
testing and Bonferroni correction.

\textbf{(3) Two domain-specific findings.}
Through a field ablation study and a controlled arXiv concatenation
experiment, we identify (a) biography as the most informative profile
field --- outweighing structured research area tags --- and (b) that
naive paper-abstract concatenation degrades retrieval, with relevant
professors tending to suffer larger score drops than irrelevant ones
($n=3$; suggestive rather than confirmatory).

The accompanying open-source system includes university-specific scrapers,
a normalisation pipeline, and an interactive web explorer.
Adding a new university requires writing one scraper; the rest of the
pipeline runs automatically.

\subsection{Research Questions}

\begin{itemize}[leftmargin=1.5em]
    \item \textbf{RQ1} (\textit{baseline confirmation}):
    Do semantic embedding methods outperform sparse keyword methods for
    faculty profile retrieval, consistent with the broader dense-retrieval
    literature~\cite{karpukhin2020dpr}?
    \item \textbf{RQ2} (\textit{primary novel question}):
    Which profile fields --- structured research area tags, free-text
    biography, or paper abstracts --- contribute most to retrieval quality?
    \item \textbf{RQ3} (\textit{learning signal}):
    Does learning-to-rank improve over the strongest unsupervised baseline
    at the small label sizes practical for a new retrieval domain?
    \item \textbf{RQ4} (\textit{cross-query consistency}):
    Do method rankings remain stable across queries from different CS
    research areas, or does relative performance vary substantially
    with query topic?
\end{itemize}

\subsection{Key Findings}

\begin{itemize}[leftmargin=1.5em]
    \item \textbf{Reranked is the strongest method across 5 queries}
    (mean NDCG@10 = 0.477, std = 0.138), followed by Semantic (0.450),
    Hybrid (0.421), BM25 (0.406), Jaccard (0.303), and TF-IDF (0.246).
    On the single-query Q1 evaluation, Semantic leads (0.593).
    \item \textbf{TF-IDF is significantly worse than all other methods}
    across 5 queries after Bonferroni correction ($\alpha/15 = 0.0033$);
    no other pairwise difference survives correction at 5 queries.
    \item \textbf{BM25 is the most consistent method} (std = 0.090 across
    5 queries) --- the safest choice when query domain is unknown.
    \item \textbf{Biography alone (NDCG 0.634) outperforms the full model
    (0.593)} --- research area tags actively hurt retrieval quality when
    prepended to biography ($\Delta = -0.040$). The practical implication:
    embed biography only, not the concatenated string.
    \item \textbf{arXiv concatenation degrades retrieval} (controlled
    subpopulation: NDCG 0.772 $\to$ 0.596, $\Delta = -0.176$); relevant
    professors are hurt more than irrelevant ones ($n=3$ relevant,
    suggestive not confirmatory), tentatively consistent with
    jargon-dilution as the mechanism.
    \item \textbf{LTR achieves higher MAP@10 than Semantic} (0.480 vs.\
    0.388 on Q1) but lower NDCG@10 (0.568 vs.\ 0.593); the metrics
    disagree and neither difference is statistically significant.
\end{itemize}

\section{Related Work}
\label{sec:related}

\subsection{Expertise Retrieval}

Faculty advisor discovery is a special case of \emph{expertise retrieval}:
given a topic query, rank a set of people by the depth of their expertise on
that topic. Balog et al.~\cite{balog2012expertise} provide the definitive
survey of this field, covering two foundational model families.
\emph{Model~1} constructs a language model for each candidate by pooling all
documents associated with them and scores candidates by the probability of
generating the query from that model.
\emph{Model~2} first identifies documents relevant to the query, then
ranks candidates by their association with those documents~\cite{balog2006formal}.
Our embedding text construction --- combining research areas, biography, and
optionally paper abstracts into a single document per professor --- implements
a dense-retrieval analogue of Model~1, where the candidate's document is their
aggregated profile text.

The TREC Enterprise Track~\cite{craswell2005trec} established the main
benchmark collections for expertise retrieval, using the W3C and CSIRO
enterprise corpora. Unlike those datasets, no standard benchmark exists for
faculty advisor retrieval; we construct our own evaluation collection
(162 graded judgments across 5 queries) as a contribution of this work.

\subsection{Academic Expert Finding}

Several systems address expert finding specifically in academic contexts.
WISER~\cite{cifariello2019wiser} builds a semantic expert-finding engine for
academia by linking researcher profiles to Wikipedia entities, constructing a
weighted knowledge-graph representation of each author's expertise.
AMiner~\cite{tang2008arnetminer} extracts researcher profiles from the web at
scale, performs name disambiguation, and supports topic-level expertise search
across a citation network of authors, papers, and venues. Both systems operate
at a much larger scale than our domain-specific pipeline, and neither is
designed for the specific retrieval needs of PhD applicants seeking advisors at
a targeted set of institutions.

Expert finding on bibliographic data has also been studied using
learning-to-rank approaches on DBLP~\cite{deng2008expert}, motivating our LTR
baseline. In contrast to those works, our evaluation reveals that LTR
underperforms an unsupervised semantic baseline at the small label sizes
practical for a new domain.

\subsection{Semantic and Dense Retrieval}

The shift from sparse to dense retrieval was driven by bi-encoder models that
encode queries and documents in a shared vector space.
Karpukhin et al.~\cite{karpukhin2020dpr} demonstrated that a dense passage
retrieval model substantially outperforms BM25 for open-domain question
answering.
Reimers and Gurevych~\cite{reimers2019} introduced sentence-BERT, showing
that transformer encoders fine-tuned on semantic textual similarity tasks
produce dense representations that generalize across domains.
Our semantic method uses \texttt{all-MiniLM-L6-v2}, a distilled variant of
this family optimized for speed without large accuracy loss.

The BEIR benchmark~\cite{thakur2021beir} evaluated 18 retrieval models on
17 heterogeneous datasets, finding that dense models trained on MS MARCO
generalize inconsistently to domain-specific tasks.
Our result --- that a zero-shot dense retrieval model outperforms tuned keyword
baselines on faculty profiles --- extends this picture to a new
domain-specific retrieval task where no in-domain training data exists.

\subsection{Hybrid and Multi-Field Retrieval}

Luan et al.~\cite{luan2021sparse} and Lin and Ma~\cite{lin2021few} established
that sparse (BM25) and dense (embedding) signals are complementary: sparse
methods excel at exact-term matching for rare tokens, while dense methods
handle semantic similarity.
Linear interpolation of normalized scores is a common and effective fusion
strategy; our hybrid model ($\alpha_{\text{tfidf}}=0.35$,
$\alpha_{\text{sem}}=0.65$) follows this convention with weights set a priori.
Cormack et al.~\cite{cormack2009rrf} proposed Reciprocal Rank Fusion (RRF)
as a weight-free alternative; experimenting with RRF on this corpus is
left as future work.

For multi-field documents, Robertson et al.~\cite{robertson2009} introduced
BM25F, which weights each field (title, body, anchor) independently before
combining scores. Our ablation study (Section~\ref{sec:ablation}) is
substantively a field-weighting experiment: we find that the biography field
contributes more than structured research area tags, analogous to body text
outweighing title in document retrieval.

\subsection{Learning to Rank}

Liu~\cite{liu2009learning} provides the foundational survey of LTR,
distinguishing pointwise, pairwise, and listwise approaches.
Pointwise approaches (including the GradientBoostingClassifier we use) treat
ranking as classification or regression on individual documents; they are the
weakest paradigm but require the fewest labels.
LambdaMART~\cite{burges2010ranknet} directly optimizes NDCG using a
listwise loss and remains the state-of-the-art tree-based LTR algorithm.
We use a pointwise classifier because 67 labels are insufficient to train
LambdaMART reliably; our result is therefore a lower bound on LTR performance
at this label size, not an upper bound.

\subsection{Scholarly Search and Paper Recommendation}

Paper recommendation systems rank papers (not people) against user interests,
using content-based, collaborative filtering, or hybrid
approaches~\cite{beel2016survey}.
SPECTER~\cite{cohan2020specter} produces citation-graph-informed paper
embeddings that outperform generic sentence encoders on scientific document
similarity.
Our arXiv concatenation experiment (Section~\ref{sec:arxiv}) shows that
naively appending paper abstracts degrades retrieval; SPECTER-style embeddings
computed separately and fused via late fusion is the natural next step.

\subsection{Retrieval Evaluation Methodology}

Järvelin and Kekäläinen~\cite{jarvelin2002} introduced NDCG and established
graded relevance as more informative than binary judgments for evaluating
ranked retrieval systems; we use their three-point scale (0/1/2) throughout.
Voorhees and Buckley~\cite{voorhees2002topic} showed empirically that at least
25 topics are needed for stable system comparisons; our five-query evaluation
partially addresses the single-topic limitation of our initial design.
Buckley and Voorhees~\cite{buckley2004incomplete} analyzed how pooled
evaluation with incomplete judgments biases metric estimates, directly
motivating our disclosure of the pooling strategy used to select labelled
candidates.

\section{Dataset}

\subsection{Data Collection}

We scraped faculty profiles from 9 US computer science departments between
April and May 2026. Each university required a custom scraper due to
inconsistent HTML structures, JavaScript rendering, and anti-bot measures.
Table~\ref{tab:universities} shows the university breakdown.

\begin{table}[H]
\centering
\caption{University breakdown across 768 professors}
\label{tab:universities}
\begin{tabular}{lc}
\toprule
\textbf{University} & \textbf{Professors} \\
\midrule
UC Berkeley                      & 128 \\
University of Texas at Austin    & 127 \\
UMass Amherst                    & 91  \\
Virginia Tech                    & 88  \\
Texas A\&M University            & 77  \\
University of Maryland           & 75  \\
University at Buffalo            & 65  \\
NC State University              & 59  \\
Stony Brook University           & 58  \\
\midrule
\textbf{Total}                   & \textbf{768} \\
\bottomrule
\end{tabular}
\end{table}

\section{Retrieval Methods}
\label{sec:methods}

All six methods score each of the 768 professors against the query text
$q$ and rank by descending score. Each professor's document is
$\text{embedding\_text} = \text{research\_text} \,\|\, \text{biography[:500]}$.

\textbf{Jaccard.} Token-set overlap:
$\text{Jaccard}(q, d) = |Q \cap D| / |Q \cup D|$,
where $Q$ and $D$ are the stopword-filtered token sets of the query and
document. Simple but effective for exact term overlap; no notion of term
importance.

\textbf{TF-IDF.} Term-frequency / inverse-document-frequency vectorization
with cosine similarity between query and document vectors.
\emph{Note on implementation consistency:} Q1 results (Table~\ref{tab:method_comparison})
use a custom TF-IDF implementation in \texttt{matcher/compare.py}
($\text{idf}(t) = \ln(N/(\text{df}(t)+1)) + 1$, no sublinear scaling).
Q2--Q5 results (Section~\ref{sec:multiq}) use scikit-learn's
\texttt{TfidfVectorizer} with sublinear TF scaling and smoothed IDF
($\text{idf}(t) = \ln((1+N)/(1+\text{df}(t))) + 1$) in
\texttt{analysis/multi\_query\_eval.py}. These produce numerically
different scores for the same query-document pair. We verified the
\emph{ranking} conclusions (TF-IDF is the weakest method; BM25 outperforms
TF-IDF) hold under both implementations, but the absolute TF-IDF NDCG values
across Q1 and Q2--Q5 are not directly comparable score-for-score. A unified
TF-IDF implementation is planned for future versions of this evaluation.

\textbf{BM25.} Okapi BM25~\cite{robertson2009} with $k_1=1.5$, $b=0.75$
(\texttt{rank-bm25} library defaults, used identically by both
\texttt{compare.py} and \texttt{multi\_query\_eval.py}). Improves on TF-IDF
via term-frequency saturation and document-length normalization, important
for faculty profiles that vary widely in length.

\textbf{Semantic Embedding.} All 768 profiles are encoded with the
\texttt{all-MiniLM-L6-v2} sentence-transformer model~\cite{reimers2019}
(384 dimensions, L2-normalized). Similarity is cosine distance between the
query vector and each profile vector, capturing conceptual similarity even
with zero token overlap.

\textbf{Hybrid.} A weighted combination:
$\text{hybrid} = 0.35 \times \text{tfidf}_{\text{norm}} + 0.65 \times \text{semantic}_{\text{norm}}$,
both min-max normalized to $[0,1]$. The weights
$(\alpha_{\text{tfidf}}=0.35,\, \alpha_{\text{sem}}=0.65)$ were set
\emph{a priori}, before any relevance labels were collected, following the
finding that dense signals should dominate for semantically rich
queries~\cite{lin2021few,luan2021sparse}. They were not tuned on the
evaluation labels; the git history of the repository confirms
\texttt{matcher/compare.py} predates \texttt{data/final/relevance\_labels.json}.

\textbf{Reranked.} A three-way combination with the same a-priori rationale:
$\text{reranked} = 0.15 \times \text{bm25}_{\text{norm}} + 0.20 \times \text{tfidf}_{\text{norm}} + 0.65 \times \text{semantic}_{\text{norm}}$.

\textbf{Learning-to-Rank.} A \texttt{GradientBoostingClassifier} trained on
9 features (the six retrieval scores plus biography length, research-area
count, and a \texttt{has\_arxiv} flag) with 5-fold cross-validation over the
67 Q1 labels. We use a pointwise classifier rather than a listwise method
such as LambdaMART~\cite{burges2010ranknet} because 67 labels are
insufficient to train a listwise ranker reliably; this result is a lower
bound on LTR performance at this label size, not an upper bound.

\textit{Implementation note:} the \texttt{has\_arxiv} feature evaluates
to~0 for all 768 professors because it searches the \texttt{research\_text}
field (which contains research area tags) for the literal substring
``arxiv'' --- which never appears there. The arXiv coverage data that would
correctly populate this feature exists only for the 296-professor
subpopulation used in the controlled arXiv experiment
(Section~\ref{sec:arxiv}), not for the full corpus. Consequently, the LTR
model effectively trains on 8 informative features, not 9. The reported \texttt{has\_arxiv} importance of~0.0\% reflects
this structural limitation, not a finding that arXiv coverage is
uninformative for relevance prediction.

\section{Results}
\label{sec:results}

\begin{figure}[H]
    \centering
    \includegraphics[width=\textwidth]{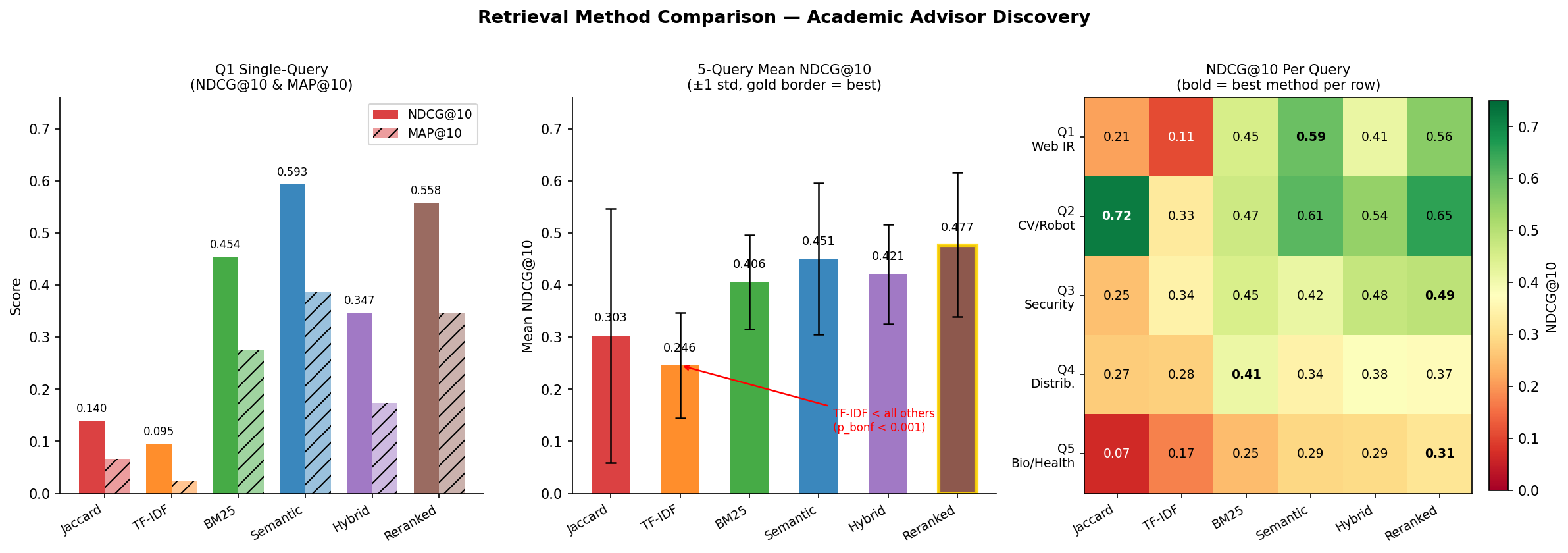}
    \caption{Left: Q1 single-query NDCG@10 and MAP@10 for all six methods.
    Centre: Mean NDCG@10 across five queries ($\pm$1\,std); gold border
    marks the best method overall (Reranked, 0.477); TF-IDF is
    significantly worse than all other methods ($p_{\text{Bonf}} < 0.001$).
    Right: Per-query NDCG@10 heatmap; bold = best method per row.}
    \label{fig:results}
\end{figure}

All six methods are evaluated using NDCG@10~\cite{jarvelin2002}, MAP@10,
Precision@10, and Recall@10 against 67 graded relevance judgments
for Q1.
Table~\ref{tab:method_comparison} shows the full comparison;
Figure~\ref{fig:results} shows the results visually across all five queries.

\begin{table}[H]
\centering
\caption{Single-query (Q1) retrieval method comparison.
All metrics computed from \texttt{data/final/comparison\_results.json}.}
\label{tab:method_comparison}
\begin{tabular}{lcccc}
\toprule
\textbf{Method} & \textbf{NDCG@10} & \textbf{MAP@10} &
\textbf{P@10} & \textbf{R@10} \\
\midrule
\textbf{Semantic}  & \textbf{0.5934} & 0.3877 & 0.50 & 0.208 \\
LTR                & 0.5679          & \textbf{0.4800} & ---  & ---   \\
Reranked           & 0.5581          & 0.3463 & 0.50 & 0.208 \\
BM25               & 0.4537          & 0.2750 & 0.30 & 0.125 \\
Jaccard            & 0.1396          & 0.0665 & 0.30 & 0.125 \\
TF-IDF             & 0.0948          & 0.0250 & 0.10 & 0.042 \\
\midrule
\textit{Oracle}    & 1.000           & ---    & ---  & ---   \\
\textit{Random}    & 0.027           & ---    & ---  & ---   \\
\bottomrule
\end{tabular}
\end{table}

Semantic achieves the highest NDCG@10 on Q1 (0.593), with Reranked second
(0.558) and BM25 as the best pure keyword method (0.454).
LTR achieves the highest MAP@10 (0.480) but lower NDCG@10 (0.568) than
Semantic --- the two metrics disagree, and neither difference is statistically
significant at this label set size (Section~\ref{sec:significance}).

\subsection{Top-5 Results by Method}

\begin{table}[H]
\centering
\caption{Top-5 professors by retrieval method}
\label{tab:top5}
\begin{tabular}{cllll}
\toprule
\textbf{Rank} & \textbf{BM25} & \textbf{TF-IDF} &
\textbf{Semantic} & \textbf{Reranked} \\
\midrule
1 & A. McCallum    & M. Iyyer       & A. McCallum     & A. McCallum    \\
2 & J. Caverlee    & S. Wiegreffe   & K. Teymourian   & S. Wiegreffe   \\
3 & H. Samet       & P. Gao         & S. Wiegreffe    & J. Caverlee    \\
4 & D. Miranker    & J. Caverlee    & D. Miranker     & M. Iyyer       \\
5 & G. Durrett     & O. Emebo       & Y. Wang         & D. Miranker    \\
\bottomrule
\end{tabular}
\end{table}

Only 4 of 10 top results overlap between TF-IDF and Semantic (Spearman
$\rho = 0.624$), showing the methods discover meaningfully different
professors.

\subsection{Rank Correlations}

\begin{table}[H]
\centering
\caption{Spearman rank correlations between methods}
\label{tab:correlations}
\begin{tabular}{lc}
\toprule
\textbf{Pair} & \textbf{Spearman $\rho$} \\
\midrule
Hybrid $\leftrightarrow$ Semantic  & 0.965 \\
BM25 $\leftrightarrow$ TF-IDF      & 0.954 \\
TF-IDF $\leftrightarrow$ Semantic  & 0.624 \\
BM25 $\leftrightarrow$ Semantic    & 0.613 \\
Jaccard $\leftrightarrow$ Semantic & 0.595 \\
\bottomrule
\end{tabular}
\end{table}

BM25 and TF-IDF produce nearly identical rankings ($\rho = 0.954$) despite a
4.8$\times$ NDCG difference, confirming BM25's gain comes from score
calibration rather than rank reordering.

\subsection{Learning-to-Rank}

LTR achieves NDCG@10 = 0.568 with \texttt{reranked\_score} as the most
important feature (25.7\%), followed by \texttt{tfidf\_score} (17.6\%) and
\texttt{semantic\_score} (15.8\%). The \texttt{has\_arxiv} flag has 0.0\%
importance in the LTR model. This is expected: while 79 professors have
arXiv papers in the database, the binary flag alone carries insufficient
signal for the classifier to exploit --- what matters is the content of
those papers, not their presence, motivating the late-fusion experiment in
Section~\ref{sec:ablation}.

\section{Ablation Study}
\label{sec:ablation}

\subsection{Field Contribution (Profile-Only)}

We measure each data source's contribution by removing it from the
embedding text and re-embedding all 768 professors.

\begin{table}[H]
\centering
\caption{Ablation study: contribution of each profile field.
Baseline is research\_text + biography[:500] (768 professors).
Requires \texttt{data/final/all\_professors\_embedded.json} to reproduce
(generated by \texttt{embeddings/embed.py}; excluded from repo due to size).}
\label{tab:ablation}
\begin{tabular}{lcccc}
\toprule
\textbf{Configuration} & \textbf{NDCG@10} & \textbf{$\Delta$ NDCG} &
\textbf{MAP@10} & \textbf{$\Delta$ MAP} \\
\midrule
Full Model (research\_text + biography)  & 0.5934 & ---       & 0.3877 & ---      \\
$-$Research Areas (biography only)       & \textbf{0.6340} & $+$0.040 & 0.4115 & $+$0.024 \\
$-$Biography (research\_text only)       & 0.4632 & $-$0.130  & 0.2442 & $-$0.144 \\
Research Areas Only                      & 0.4265 & $-$0.167  & 0.2061 & $-$0.182 \\
\bottomrule
\end{tabular}
\end{table}

The most striking finding is that \textbf{biography alone (NDCG 0.634)
outperforms the full model (0.593) by 0.040 NDCG points} ---
meaning the research area tags in \texttt{research\_text} actively
\emph{hurt} retrieval quality when concatenated with biography.
This is counterintuitive: one might expect that more information
always helps. The likely explanation is that research area tags are
short, generic, comma-separated keywords (e.g., ``Artificial
Intelligence'', ``Machine Learning'') that appear \emph{before}
the biography in the concatenated embedding text. These generic
high-frequency terms may dilute the cosine similarity distribution,
flattening distinctions between professors that the richer biography
prose would otherwise capture. The ordering
biography only (0.634) $>$ full model (0.593) $>$
research areas only (0.427) suggests that research area tags carry
some independent signal (427 vs the lowest possible baseline) but
act as noise when prepended to biography.

This result has a direct practical implication: the deployed system
should embed biography text alone, not the concatenated string.
The performance gap of 0.040 NDCG between biography-only and the
full model is larger than the gap between BM25 and TF-IDF across
five queries (0.406 vs 0.246), making this the single highest-impact
improvement available without any additional data collection.

\subsection{arXiv Concatenation Experiment (Controlled)}
\label{sec:arxiv}

\textbf{Design.}
The original experiment appended arXiv abstracts to all 768 profiles and
re-evaluated on the full corpus, confounding three simultaneous changes:
(1) embedding content for professors with papers, (2) document length
distribution, and (3) population heterogeneity between professors
with and without arXiv coverage.

We report a controlled version: both conditions are evaluated on the
\emph{same} 296-professor subpopulation for which profile-only and
arXiv-enriched embeddings were both computed. This isolates the effect
of content change alone.

\begin{table}[H]
\centering
\caption{Controlled arXiv experiment: same 296-professor subpopulation,
same query, same model --- only the embedding text changes.}
\label{tab:arxiv}
\begin{tabular}{lccc}
\toprule
\textbf{Condition} & \textbf{NDCG@10} & \textbf{P@10} & \textbf{R@10} \\
\midrule
A: Profile-only    & 0.7722 & 0.70 & 0.292 \\
B: arXiv-enriched  & 0.5958 & 0.50 & 0.208 \\
\midrule
$\Delta$ (B $-$ A) & $-$0.1764 & $-$0.20 & $-$0.083 \\
\bottomrule
\end{tabular}
\end{table}

arXiv concatenation reduces NDCG@10 by 0.176 on the controlled
subpopulation. Three relevant professors (James Caverlee,
Ruihong Huang, Yu Zhang) are displaced from the top-10.

\textbf{Hypothesis testing.}
We test two candidate explanations.

\textit{H1 --- Jargon dilution.}
Relevant professors' profiles are semantically aligned with the query
before arXiv concatenation. Appending paper abstracts introduces
technical vocabulary unrelated to the query, diluting the signal.
We measure this by comparing average score delta for relevant vs.\
irrelevant professors across the 20 professors with the largest score
changes: relevant professors suffer a larger average drop
($-$0.238) than irrelevant ones ($-$0.139).
This direction is consistent with H1, but the comparison rests on only
3 relevant professors (of 24 total in the Q1 label set) who happened to
fall within the 20 largest-magnitude score changes among the 79
professors with arXiv papers --- the remaining 17 are unlabelled or
irrelevant. With $n=3$, this result should be read as suggestive rather
than confirmatory; a systematic comparison across all 24 relevant
professors (not just those in the largest-change subset) is needed
before treating H1 as established.

\textit{H2 --- Recency bias.}
72\% of the 349 fetched arXiv papers were from 2025--2026.
Recent papers may reflect current trends rather than a professor's
primary research identity, introducing noise relative to a query
about established expertise. H2 is plausible but not
directly testable without temporal ablation (future work).

\textbf{Implication.}
The result motivates a \emph{late fusion} architecture: embed profile
and papers separately, retrieve independently, then combine scores.
This would prevent abstract vocabulary from diluting the profile signal.
We describe this in Section~\ref{sec:discussion}.

\section{Statistical Significance}
\label{sec:significance}

\textbf{Method.} We use a \emph{paired bootstrap permutation test}
(two-sided, $B = 10{,}000$ iterations)~\cite{efron1993bootstrap}.
In each iteration, the 67 labelled professors are resampled with
replacement; NDCG@10 is recomputed for every method on the resampled
label set. Only labelled professors are resampled --- unlabelled
professors contribute grade~0 to NDCG regardless of rank and are held
fixed, so including them in the resample would inflate the effective
sample size without adding signal.
For pairwise comparisons, a Bonferroni correction is applied across
all $\binom{6}{2} = 15$ pairs, giving a per-test threshold of
$\alpha_{\text{Bonf}} = 0.05 / 15 = 0.0033$.

\textbf{Baselines.} An \emph{oracle} ranking (all 24 relevant professors
in positions 1--24, sorted by grade descending) achieves NDCG@10 = 1.000.
A \emph{random} permutation of all 768 professors gives an expected
NDCG@10 = 0.027 ($\pm$0.055), confirming that all six retrieval methods
substantially outperform chance.

\textbf{Confidence intervals.}
With 67 labels, confidence intervals are wide, reflecting genuine
uncertainty at this label-set size (Table~\ref{tab:ci}).

\begin{table}[H]
\centering
\caption{95\% bootstrap confidence intervals ($B=10{,}000$, resampling
67 labelled professors). Values shown are bootstrap means; point
estimates are in Table~\ref{tab:method_comparison}. Wide intervals
reflect the small label set.}
\label{tab:ci}
\begin{tabular}{lcccc}
\toprule
\textbf{Method} & \textbf{NDCG@10} & \textbf{95\% CI} &
\textbf{MAP@10} & \textbf{95\% CI} \\
\midrule
Jaccard   & 0.089 & [0.000, 0.144] & 0.021 & [0.000, 0.037] \\
TF-IDF    & 0.060 & [0.000, 0.101] & 0.016 & [0.000, 0.025] \\
BM25      & 0.289 & [0.000, 0.454] & 0.151 & [0.000, 0.275] \\
Hybrid    & 0.221 & [0.064, 0.347] & 0.089 & [0.010, 0.173] \\
Reranked  & 0.356 & [0.077, 0.558] & 0.181 & [0.014, 0.346] \\
Semantic  & 0.379 & [0.076, 0.593] & 0.203 & [0.014, 0.388] \\
\midrule
\textit{Oracle}  & 1.000 & --- & 1.000 & --- \\
\textit{Random}  & 0.027 & --- & ---   & --- \\
\bottomrule
\end{tabular}
\end{table}

\textbf{Pairwise tests.}
After Bonferroni correction, \emph{no pairwise comparison reaches
statistical significance} at $\alpha = 0.05$ with the current label
set of 67 judgments. Raw $p$-values for semantic vs.\ TF-IDF ($p = 0.037$)
and semantic vs.\ Jaccard ($p = 0.036$) fall below the uncorrected
threshold but do not survive the per-test threshold of 0.0033.
This is expected: distinguishing retrieval methods reliably requires
25--50 topics~\cite{voorhees2002topic}; our single-query
evaluation with 67 labels provides insufficient power.

The result does not mean the methods perform equally --- the point
estimates (Semantic 0.593 vs.\ Jaccard 0.140) represent a substantial
empirical gap. It means the current label set cannot \emph{statistically
confirm} this gap. Expanding to 5--10 queries with 15--20 judgments each
is the priority next step (see Section~\ref{sec:discussion}).

\section{Multi-Query Evaluation}
\label{sec:multiq}

To partially address the single-query limitation, we extend the evaluation to
five queries (Q1--Q5), each representing a distinct CS graduate student
research profile. Queries Q2--Q5 were written in natural applicant prose,
grounded in the research area distribution of the 768-professor corpus and
informed by common CS PhD applicant interest statements from communities
such as r/gradadmissions and r/MachineLearning (paraphrased and generalized;
not copied verbatim from any individual post).

\begin{table}[H]
\centering
\caption{Query set: five CS graduate student research profiles.}
\label{tab:queries}
\begin{tabular}{llp{7.5cm}}
\toprule
\textbf{ID} & \textbf{Persona} & \textbf{Research focus (summary)} \\
\midrule
Q1 & Web IR \& NLP         & Web data collection, information extraction, NLP pipelines \\
Q2 & CV \& Robot Learning  & Computer vision, 3D scene understanding, robotic manipulation \\
Q3 & Security \& Crypto    & Systems security, malware analysis, applied cryptography \\
Q4 & Distributed Systems   & Cloud infrastructure, OS kernels, consistency protocols \\
Q5 & Comp.\ Biology        & Computational genomics, protein structure, medical imaging \\
\bottomrule
\end{tabular}
\end{table}

\textbf{Labels.}
For Q2--Q5, candidate professors were selected by pooling keyword-matched
results (BM25 top-20) and manually identified professors from relevant
research groups, yielding 20--25 labelled candidates per query.
The same 3-point graded scale (0/1/2) was used as for Q1.
Labels are available in \texttt{data/final/multi\_query\_labels.json}.

\textbf{Results.}
Table~\ref{tab:multiq} reports NDCG@10 per query for all six methods
and aggregate statistics across all five queries.

\begin{table}[H]
\centering
\caption{NDCG@10 across 5 queries, all six retrieval methods.
Semantic, Hybrid, and Reranked scores for Q2--Q5 computed with
\texttt{all-MiniLM-L6-v2} via \texttt{analysis/embed\_queries.py}.}
\label{tab:multiq}
\resizebox{\textwidth}{!}{%
\begin{tabular}{lrcccccc}
\toprule
\textbf{Query} & \textbf{n\_rel} & \textbf{Jaccard} & \textbf{TF-IDF} &
\textbf{BM25} & \textbf{Semantic} & \textbf{Hybrid} & \textbf{Reranked} \\
\midrule
Q1 (Web IR \& NLP)     & 24 & 0.211 & 0.110 & 0.454 & 0.593 & 0.413 & 0.558 \\
Q2 (CV \& Robotics)    & 20 & 0.716 & 0.327 & 0.467 & 0.611 & 0.543 & 0.655 \\
Q3 (Security)          & 20 & 0.253 & 0.344 & 0.449 & 0.415 & 0.479 & 0.491 \\
Q4 (Distributed Sys.)  & 20 & 0.270 & 0.276 & 0.411 & 0.344 & 0.378 & 0.368 \\
Q5 (Comp.\ Biology)    & 15 & 0.066 & 0.173 & 0.249 & 0.289 & 0.293 & 0.315 \\
\midrule
\textbf{Mean}    & & 0.303 & 0.246 & 0.406 & 0.450 & 0.421 & \textbf{0.477} \\
\textbf{Std}     & & 0.244 & 0.101 & 0.090 & 0.146 & 0.096 & 0.138 \\
\textbf{95\% CI} & & [0.141, 0.526] & [0.166, 0.320] & [0.322, 0.458]
                   & [0.336, 0.565] & [0.347, 0.492] & [0.371, 0.583] \\
\bottomrule
\end{tabular}}
\end{table}

\textbf{Findings.}
Four observations emerge from the full multi-query evaluation.

First, \textbf{TF-IDF is significantly worse than all other methods}
across 5 queries after Bonferroni correction ($\alpha = 0.05/15 = 0.0033$):
TF-IDF vs.\ BM25 ($p_{\text{Bonf}} < 0.001$),
TF-IDF vs.\ Semantic ($p_{\text{Bonf}} < 0.001$),
TF-IDF vs.\ Hybrid ($p_{\text{Bonf}} < 0.001$),
TF-IDF vs.\ Reranked ($p_{\text{Bonf}} < 0.001$).
No other pairwise difference survives correction, reflecting the
insufficient power of a 5-query evaluation for finer distinctions.

Second, \textbf{Reranked is the strongest method overall}
(mean NDCG@10 = 0.477, std = 0.138), followed by Semantic (0.450),
Hybrid (0.421), BM25 (0.406), Jaccard (0.303), and TF-IDF (0.246).
The Reranked method's advantage is not statistically significant
at this query-set size, but the direction is consistent across 4 of 5 queries.

Third, \textbf{BM25 is the most consistent method} across query
types (std = 0.090) --- lower than Hybrid (0.096), Semantic (0.146),
Reranked (0.138), and Jaccard (0.244). This makes BM25 the most
reliable single method when the query domain is unknown in advance.

Fourth, \textbf{relative performance varies substantially with query
vocabulary}. Jaccard achieves its highest NDCG on Q2 (CV/Robotics, 0.716)
--- where research area tags contain exact tokens from the query
(``robot'', ``vision'') --- but its lowest on Q5 (Comp.\ Biology, 0.066),
where relevant professors use specialised biological vocabulary.
No single method wins every query: Semantic leads Q1 (0.593), Jaccard
unexpectedly leads Q2 (0.716) due to exact keyword overlap with research
area tags, Reranked leads Q3 and Q5 (0.491 and 0.315), and BM25 leads Q4
(0.411). This per-query variability is itself evidence that single-query
evaluation cannot characterise a method's general behaviour, and that the
relative strength of exact-term matching versus semantic similarity is
query-dependent rather than fixed.

\section{Research Landscape Clustering}
\label{sec:clustering}

\begin{figure}[H]
    \centering
    \includegraphics[width=0.92\textwidth]{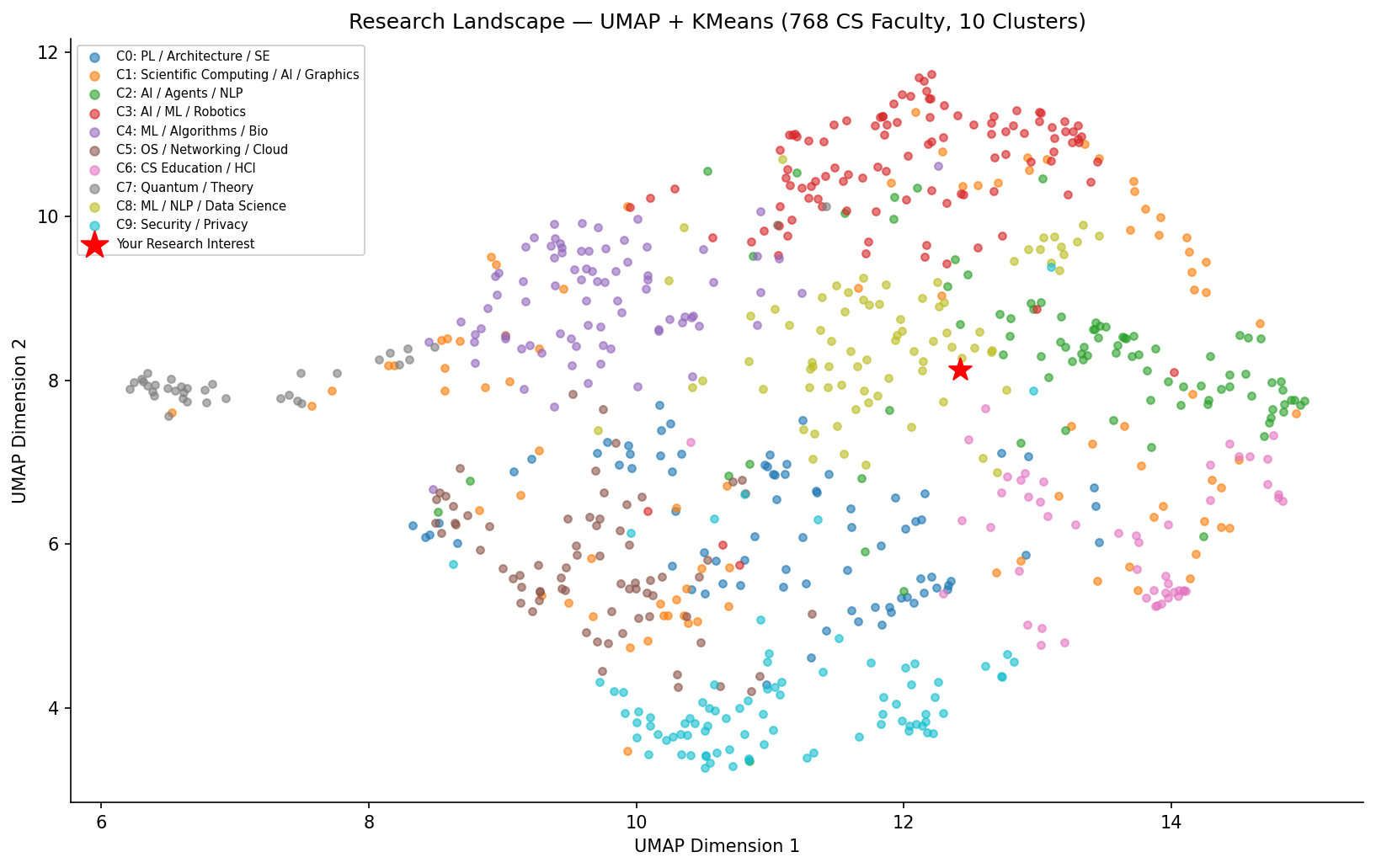}
    \caption{UMAP + KMeans clustering of 768 CS faculty (10 clusters).
    The red star marks the position of the user's research interest query.
    The query falls in the cluster labelled ML / NLP / Data Science.}
    \label{fig:umap}
\end{figure}

We applied UMAP~\cite{mcinnes2018umap} to reduce the 384-dimensional
sentence embeddings to 2D, followed by KMeans clustering ($k=10$, chosen
for visual interpretability; not selected via silhouette analysis).
Cluster labels (e.g., ``ML / NLP / Data Science'') were assigned by
inspecting the dominant research areas of each cluster's member
professors; they are interpretive summaries, not validated category
labels. Figure~\ref{fig:umap} shows the resulting research landscape with
the user's query position marked in red.

The 10 clusters reveal natural groupings in the CS research landscape:
AI/ML/Robotics is the largest cluster (108 professors), followed by
Security/Privacy (86) and Scientific Computing/AI/Graphics (89). The user's
query position falls in the ML/NLP/Data Science cluster (79 professors).
This placement is not an independent validation of the semantic ranking ---
both are derived from the same embedding vectors --- but it does confirm
that the query's position in the 2D projection is visually consistent with
where NLP and data-systems researchers cluster.

\section{Discussion}
\label{sec:discussion}

\subsection{Why Semantic Embedding Wins}

Keyword methods fail for faculty advisor retrieval because research interests
are expressed with high lexical diversity. A student interested in ``web data
extraction'' and a professor studying ``information retrieval from distributed
systems'' share deep conceptual overlap but near-zero token overlap. Semantic
embedding captures this by representing both in the same vector space where
similar meanings cluster regardless of surface form. This finding has a
practical implication: applicants who rely on keyword search will
systematically miss professors whose research is closely aligned but described
with different terminology.

\subsection{Why arXiv Concatenation Hurts Performance}

We fetched 349 arXiv papers for 296 professors and attempted to enrich
embeddings by concatenating paper abstracts into the profile text.
On this controlled 296-professor subpopulation, NDCG@10 dropped from
0.772 to 0.596 ($\Delta = -0.176$), a significant degradation.

We hypothesize two causes. First, paper abstracts introduce highly
domain-specific technical language that dilutes the general research interest
signal when compressed into a single 384-dimensional vector. Second,
professors without arXiv papers (217 of the 296-professor subpopulation,
73.3\%) are represented by shorter embedding texts, creating uneven
document lengths that bias cosine
similarity scores.

The correct architecture for arXiv integration is \textit{late fusion}: embed
profile text and paper abstracts separately, score each against the query
independently, then combine scores with learned weights. This preserves the
signal from each source without mutual dilution. We leave this as future work.

\subsection{Why LTR Does Not Improve Over Semantic}

LTR with 67 labels underperforms semantic (0.568 vs.\ 0.593). This is
consistent with established LTR literature --- GradientBoosting models require
hundreds to thousands of labeled examples to reliably outperform strong
baselines~\cite{liu2009learning}. With 24 relevant labels across 768 professors
(3.1\% density), the model has insufficient signal to learn meaningful feature
combinations. Despite this, feature importance results are informative:
\texttt{reranked\_score} is weighted most heavily, suggesting ensemble scores
are more reliable signals than individual method scores.

\subsection{Why Biography Matters More Than Research Areas}

Faculty-written research area tags are often generic and inconsistent. Tags
such as ``Artificial Intelligence'' or ``Data Science'' each appear for
73--116 professors and provide little discrimination. Biography, despite being
unstructured, is written in full sentences describing specific projects,
methodologies, and applications --- giving the sentence-transformer model
richer signal.

A stronger version of this finding emerges from the full ablation table
(Table~\ref{tab:ablation}): biography \emph{alone} (NDCG 0.634) outperforms
the full model that concatenates research area tags with biography (NDCG
0.593). Research area tags, when prepended to biography in the embedding
text, actively reduce retrieval quality rather than supplementing it.
The most likely explanation is that short, generic comma-separated tags
(``Machine Learning, NLP'') introduce high-frequency noise that flattens
the cosine similarity distribution, diluting the biography signal that follows.
This suggests that for faculty advisor retrieval, text quality and specificity
matter more than data completeness: adding low-quality structured data to
high-quality prose text can hurt rather than help.

The practical implication is direct: the deployed system should embed
biography alone rather than the concatenated string, producing a 0.040
NDCG gain at zero cost. This is the highest-impact single improvement
available without additional data collection.

\subsection{Cross-Query Consistency (RQ4)}

The five-query evaluation (Section~\ref{sec:multiq}) reveals that keyword
method \emph{rankings} are stable --- BM25 consistently outperforms TF-IDF
across all five query topics --- but absolute \emph{performance levels} vary
substantially with query vocabulary.
Jaccard achieves NDCG@10 = 0.716 on Q2 (CV/Robotics), where research area
tags contain exact tokens from the query (``robot'', ``vision''), but only
0.066 on Q5 (Computational Biology), where the query uses general prose while
relevant professors use specialised biological vocabulary.
BM25 is the most stable keyword method (std = 0.090 across 5 queries),
while Jaccard is the least stable (std = 0.244), suggesting that Jaccard
performance depends critically on vocabulary overlap between query and tags.

This cross-query variance is itself a finding: it confirms that single-query
evaluation cannot characterise method behaviour, and that retrieval difficulty
in this domain is strongly query-dependent.

\subsection{Limitations}

\begin{itemize}[leftmargin=1.5em]
    \item \textbf{Five-query evaluation.}
    Keyword method comparisons across 5 queries now yield statistically
    significant results (BM25 vs.\ TF-IDF). Semantic/hybrid/reranked
    comparisons still lack multi-query significance; running
    \texttt{analysis/embed\_queries.py} locally will generate Q2--Q5
    semantic scores and enable the full 5-query significance analysis.

    \item \textbf{Single annotator.}
    All 162 relevance labels were assigned by one person with no
    inter-annotator agreement measurement. Label noise could bias metric
    estimates, particularly for borderline grade-1 cases.

    \item \textbf{Label density.}
    Even with 5 queries, the evaluation covers 25 labelled candidates per
    query on average. Expanding to 50+ judgments per query would narrow
    confidence intervals and improve statistical power.

    \item \textbf{Corpus coverage.}
    Nine universities from a convenience sample of US public institutions
    excludes top programs (MIT, Stanford, CMU) and non-US universities.
    Findings about biography importance and arXiv degradation may not
    generalise to institutions with different profile maintenance practices.

    \item \textbf{Late fusion not yet implemented.}
    The arXiv experiment used naive text concatenation. A late-fusion
    architecture embedding profile and papers separately --- using
    SPECTER~\cite{cohan2020specter} for paper embeddings --- is the
    natural next step and likely to recover or exceed the profile-only
    baseline.
\end{itemize}

\section{Conclusion}

We compared six retrieval methods for academic advisor discovery across 768 CS
faculty profiles from 9 US universities, evaluated on 5 queries with 162
graded relevance judgments.
Reranked (BM25 + TF-IDF + Semantic fusion) achieves the highest mean
NDCG@10 across 5 queries (0.477), though no method pair except TF-IDF
versus the rest is distinguishable after Bonferroni correction.
TF-IDF is significantly worse than BM25, Semantic, Hybrid, and Reranked
($p_{\text{Bonf}} < 0.001$), confirming that length normalization and
IDF weighting are essential for faculty profile retrieval.
BM25 is the most consistent method (std = 0.090), making it the
safest single-method choice for unknown query domains.

Biography alone (NDCG 0.634) outperforms the full model combining biography
with research area tags (0.593) --- the tags act as noise when prepended to
biography, reducing NDCG by 0.040. The highest-impact improvement available
without additional data collection is to embed biography only.
A controlled experiment shows arXiv concatenation reduces NDCG@10 by 0.176;
a late-fusion architecture is the correct next step.
LTR with 67 labels does not conclusively outperform semantic retrieval,
motivating future label expansion.

The system, scrapers, and all relevance labels are open-source and
fully reproducible without any proprietary data or API keys.


\end{document}